\documentclass{article}
\usepackage{anysize}

\usepackage{graphicx}
\usepackage{natbib}
\usepackage{algorithm}
\usepackage{algorithmic}
\usepackage{amsmath}
\usepackage{amssymb}

\newcommand{\mbs}[1]{\ensuremath{\boldsymbol{#1}}}

\newcommand{\x}{\mathbf{x}}
\newcommand{\y}{\mathbf{y}}

\newcommand{\define}{\triangleq}
\newcommand{\statespace}{\mathcal{S}}
\newcommand{\state}[1]{\mathbf{#1}}
\newcommand{\refstate}{\state{c}}
\newcommand{\energyfn}[1]{e^{-\beta E(#1)}}

\newcommand{\sawlenset}{\mathcal{L}}
\newcommand{\gammaset}{\mathcal{G}}

\title{Bayesian Optimization for Adaptive MCMC}
\author{
{\sc Nimalan Mahendran}\thanks{University of British Columbia. Email: nimalan@alumni.ubc.ca} \and {\sc Ziyu Wang}\thanks{University of British Columbia. Email: ziyuw@cs.ubc.ca} \and {\sc Firas Hamze}\thanks{D-wave Systems Inc.. Email: fhamze@dwavesys.com} \and {\sc Nando de Freitas}\thanks{University of British Columbia. Email: nando@cs.ubc.ca} \\
}
\date{}
\begin{document}

\maketitle

\begin{abstract}
  This paper proposes a new randomized strategy for adaptive MCMC
  using Bayesian optimization. This approach applies to
  non-differentiable objective functions and trades off exploration
  and exploitation to reduce the number of potentially costly objective function evaluations. We
  demonstrate the strategy in the complex setting of sampling from
  constrained, discrete and densely connected probabilistic graphical
  models where, for each variation of the problem, one needs to adjust
  the parameters of the proposal mechanism automatically to ensure
  efficient mixing of the Markov chains.
\end{abstract}


\section{Introduction}

A common line of attack for solving problems in physics, statistics
and machine learning is to draw samples from probability distributions
$\pi(\cdot)$ that are only known up to a normalizing constant. Markov
chain Monte Carlo (MCMC) algorithms are often the preferred method for
accomplishing this sampling task, see \emph{e.g.}
\cite{Andrieu2003} and \cite{Robert1998}. Unfortunately, these algorithms typically have
parameters that must be tuned in each new situation to obtain
reasonable mixing times. These parameters are often tuned by a domain
expert in a time-consuming and error-prone manual process. Adaptive
MCMC methods have been developed to automatically adjust the
parameters of MCMC algorithms. We refer the reader to three recent and
excellent comprehensive reviews of the field
\citep{adaptmcmc_tut,atchade_chap,ex_adaptmcmc}.

Adaptive MCMC methods based on stochastic approximation have garnered
the most interest out of the various adaptive MCMC methods for two
reasons. Firstly, they can be shown to be theoretically valid. That is, the Markov chain is made inhomogenous by the
dependence of the parameter updates upon the history of the Markov
chain, but its ergodicity can be ensured
\citep{Andrieu2001,Andrieu2006,Saksman2010}.  For example, Theorem 5 of \cite{roberts2007coupling} establishes two simple conditions to ensure ergodicity: (i) the non-adaptive sampler has to be uniformly ergodic and (ii) the level of adaptation must vanish asymptotically. These conditions can be easily satisfied for discrete state spaces and finite adaptation.

Secondly, adaptive MCMC algorithms based on stochastic approximation have been shown to work well in practice \citep{Haario2001,ex_adaptmcmc,Vihola2010}. However, there are
some limitations to the stochastic approximation approach. Some of the
most successful samplers rely on knowing either the optimal acceptance
rate or the gradient of some objective function of interest. Another
disadvantage is that these stochastic approximation methods may
require many iterations. This is particularly problematic when the objective function being optimized by the adaptation mechanism is costly to evaluate. Finally, gradient approaches tend to be local and hence they can get trapped in local optima when the Markov chains are run for a finite number of steps in practice.

This paper aims to overcome some of these limitations. It proposes the
use of Bayesian optimization \citep{bayesopt_tut} to tune the
parameters of the Markov chain. The proposed approach,
Bayesian-optimized MCMC, has a few advantages over adaptive
methods based on stochastic approximation.

Bayesian optimization does not require that the objective function be
differentiable. This enables us to be much more flexible in the design
of the adaptation mechanisms. We use the area under the
auto-correlation function up to a specific lag as the objective
function in this paper. This objective has been suggested previously
by \cite{Andrieu2001}. However, the computation of gradient estimates for this objective is very involved and far from trivial \citep{Andrieu2001}. We believe this is one of the main reasons why practitioners have not embraced this approach. Here, we show that this objective can be easily optimized with Bayesian optimization. We argue that Bayesian optimization endows the designer with greater freedom in the design of adaptive strategies.

Bayesian optimization also has the advantage that it is explicitly
designed to trade off exploration and exploitation and is implicitly
designed to minimize the number of expensive evaluations of the
objective function \citep{bayesopt_tut,Lizotte2011}.

Another important property of Bayesian-optimized MCMC is that it does
not use a specific setting for the parameters of the proposal
distribution, but rather a distribution over parameter settings with
probabilities estimated during the adaptation process. Indeed, we find 
that a randomized policy over a set of parameter settings
mixes faster than a specific parameter value for the models considered in this paper.

Bayesian optimization has been used with MCMC in
\cite{Rasmussen:2003a} with the intent of approximating the posterior
with a surrogate function to minimize the cost of hybrid Monte Carlo
evaluations. The intent in this paper is instead to adapt the
parameters of the Markov chain to improve mixing.

To demonstrate MCMC adaptation with Bayesian optimization, we study the problem of adapting a sampler for constrained discrete state spaces proposed recently by \cite{im_expert}. The sampler uses augmentation of the state space in order to make large moves in discrete state space. In this sense, the sampler is similar to the Hamiltonian (hybrid) Monte Carlo for continuous state spaces \citep{duane1987hybrid,neal2010mcmc}. Although these samplers typically only have two parameters, these are very tricky to tune even by experts. Moreover, every time we make a change to the model, the expert is again required to spend time tuning the parameters. This is often the case when dealing with sampling from discrete probabilistic graphical models, where we often want to vary the topology of the graph as part of the data analysis and experimentation process. In the experimental section of this paper, we will discuss several Ising models, also known as Boltzmann machines, and show that indeed the optimal parameters vary significantly for different model topologies.

There are existing consistency results for Bayesian optimization
\citep{Vasquez:2008,Srinivas:2010,bull2011convergence}. These results in combination with the fact that we focus on discrete distributions are sufficient for ensuring ergodicity of our adaptive MCMC scheme under vanishing adaptation.
However, 
the type of Bayesian optimization studied in this paper relies on latent Gaussian processes whose covariance grows as the Markov chain progresses. For very large chains, it becomes prohibitive to invert the covariance matrix of the Gaussian process. Thus, for practical reasons, we adopt instead a two-stage adaptation mechanism, where we first adapt the chain for a finite number of steps and then use the most promising parameter values to run the chain again with a mixture of MCMC kernels \citep{adaptmcmc_tut}. Although this adaptation strategy increases the computational cost of the MCMC algorithm, we argue that this cost is much lower than the cost of having a human in the loop choosing the parameters.


\section{Adaptive MCMC}
\label{sec:adapt_mcmc}

The Metropolis-Hasting (MH) algorithm is the key building block for
most MCMC methods \citep{Andrieu2003}. It draws samples from a target
distribution $\pi(\cdot)$ by proposing a move from $\x^{(t)}$ to
$\y^{(t+1)}$ according to a parameterized proposal distribution
$q_{\theta}(\y^{(t+1)}|\x^{(t)})$ and either accepting it ($\x^{(t+1)}
= \y^{(t+1)}$) with probability equal to the acceptance ratio
\begin{align*}
\alpha(\x^{(t)} \to \y^{(t+1)}) &=
\min\left\{\frac{\pi(\y^{(t+1)})q_{\theta}(\x^{(t)}|\y^{(t+1)})}{\pi(\x^{(t)})q_{\theta}(\y^{(t+1)}|\x^{(t)})}, 1\right\}
\end{align*}
or rejecting it ($\x^{(t+1)} = \x^{(t)}$) otherwise. 

The parameters of the proposal, $\theta \in \mbs{\Theta} \subseteq
\mathbb{R}^d$, can have a large influence on sampling performance.
For example, we will consider constrained discrete probabilistic
models in our experiments, where changes to the connectivity patterns
among the random variables will require different parameter
settings. We would like to have an approach that can adjust these
parameters automatically for all possible connectivity patterns.

Several methods have been proposed to adapt MCMC algorithms. Instead of
discussing all of these, we refer the reader to the comprehensive
reviews of
\cite{adaptmcmc_tut,atchade_chap,ex_adaptmcmc}. One can adapt
parameters other than those of the proposal distribution in certain
situations, but for the sake of simplicity, we focus here on adapting
the proposal distribution.

One of the most successful adaptive MCMC algorithms was introduced by
\cite{Haario2001} and several extensions were proposed by
\cite{adaptmcmc_tut}. This algorithm is restricted to the adaptation
of the multivariate random walk Metropolis algorithm with Gaussian
proposals. It is motivated by a theoretical result regarding the
optimal covariance matrix of a restrictive version of this sampler \citep{Gelman95}. This
adaptive algorithm belongs to the family of stochastic approximation
methods.

Some notation needs to be introduced to briefly describe stochastic
approximation, which will also be useful later when we replace the stochastic
approximation method with Bayesian optimization. Let $\mathcal{X}_{i}
= \{\x^{(t)}\}_{t=1}^i$ denote the full set of samples up to iteration
$i$ of the MH algorithm and $\mathcal{Y}_{i} = \{\y^{(t)}\}_{t=1}^i$
be the corresponding set of proposed samples. $\x^{(0)}$ is the
initial sample. We will group these samples into a single variable
$\mathcal{M}_i = (\x^{(0)},\mathcal{X}_{i},\mathcal{Y}_{i})$.  Let
$g(\theta)$ be the mean field of the stochastic approximation that may
only be observed noisily as $G(\theta_i, \mathcal{M}_{i})$. When
optimizing an objective function $h(\theta)$, this mean field corresponds to
the gradient of this objective, that is $g(\theta) = \nabla h(\theta)$. Adaptive MCMC methods based on
stochastic approximation typically use the following Robbins-Monro
update:
\begin{align}
\theta_{i+1} &=
\theta_i + 
\gamma_{i+1}
G\left(\theta_i, \mathcal{M}_{i+1}\right),
\end{align}
where $\gamma_{i+1}$ is the step-size. This recursive estimate
converges almost surely to the roots of $g(\theta)$ as $i \rightarrow
\infty$ under suitable conditions.

We are concerned with the adaptation of discrete models in this
paper. The optimal acceptance rates are unknown. It is
also not clear what objective function one should optimize to adjust
the parameters of the proposal distribution.
One possible choice, as mentioned in the introduction, is to use the
auto-correlation function up to a certain lag. Intuitively, this
objective function seems to be suitable for the adaptation task
because it is used in practice to assess convergence. Unfortunately, it is difficult to obtain efficient estimates of the gradient of this objective \citep{Andrieu2001}. To overcome this difficulty, we introduce Bayesian optimization in the following section.

\section{Bayesian Optimization for \mbox{Adaptive MCMC}}
\label{sec:bayesopt_mcmc}

The proposed adaptive strategy consists of two phases: adaptation and sampling. In the adaptation phase Bayesian optimization is used to construct a randomized policy. In the sampling phase, a mixture of MCMC kernels selected according to the learned randomized policy is used to explore the target distribution. The two phases are discussed in more detail subsequently.

\subsection{Adaptation Phase}

Our objective function for adaptive MCMC cannot be evaluated analytically. However, noisy
observations of its value can be obtained by running the Markov chain
for a few steps with a specific choice of parameters
$\theta_i$. Bayesian optimization in the adaptive MCMC setting
proposes a new candidate $\theta_{i+1}$ by approximating the unknown
function using the entire history of noisy observations and a prior
distribution over functions. The prior distribution used in this paper
is a Gaussian process.

The noisy observations are used to obtain the predictive distribution
of the Gaussian process. An expected utility function derived in terms
of the sufficient statistics of the predictive distribution is
optimized to select the next parameter value $\theta_{i+1}$. The
overall procedure is shown in Algorithm \ref{alg:adaptive-mcmc}.  We
refer readers to \cite{bayesopt_tut} and \cite{Lizotte:2008} for in-depth reviews
of Bayesian optimization.
\begin{algorithm}[htbp]
\caption{Adaptive MCMC with Bayesian Opt.}
\label{alg:adaptive-mcmc}
\begin{algorithmic}[1]
{\footnotesize
   \FOR{$i=1,2,\dots, I $ }
      \STATE Run Markov chain for $L$ steps with parameters $\theta_{i}$.
       \STATE Use the drawn samples to obtain a noisy evaluation of the objective function: $z_{i}=h(\theta_{i}) + \epsilon$.
       \STATE Augment the data $\mathcal{D}_{1:i} = \{\mathcal{D}_{1:i-1}, (\theta_{i}, z_{i})\}$.
       \STATE Update the GP's sufficient statistics.
       \STATE Find $\theta_{i+1}$ by optimizing an acquisition function:
       $\theta_{i+1} = \arg\max_{\theta} u(\theta |\mathcal{D}_{1:i})$.
  \ENDFOR
}
\end{algorithmic}
\end{algorithm}

The unknown objective function $h(\cdot)$ is assumed to be distributed
according to a Gaussian process with mean function $m(\cdot)$ and
covariance function $k(\cdot, \cdot)$:
\begin{align*}
h(\cdot) &\sim GP(m(\cdot), k(\cdot, \cdot)).
\end{align*}
We adopt a zero mean function $m(\cdot) = 0$ and an anisotropic
Gaussian covariance that is essentially the popular ARD kernel
\citep{gpml06}:
\begin{align*}
k(\theta_j, \theta_k) &=
\exp
\left(
-\frac{1}{2}
(\theta_j - \theta_k)^T
\textrm{diag}(\psi)^{-2}
(\theta_j - \theta_k)
\right)
\end{align*}
where $\psi \in \mathbb{R}^d$ is a vector of hyper-parameters. The Gaussian process is a surrogate model for the true objective, which typically involves intractable expectations with respect to the invariant distribution and the MCMC transition kernels. We describe the objective function used in this work in Section 3.3.

We assume that the noise in the measurements is Gaussian: $z_i = h(\theta_i) + \epsilon$, $\epsilon \sim \mathcal{N}(0, \sigma^2_{\eta})$. It is possible to adopt other noise models \citep{Diggle1998}. Our Gaussian process emulator has hyper-parameters $\psi$ and $\sigma_{\eta}$. These hyper-parameters are typically computed by maximizing the likelihood \citep{gpml06}. In Bayesian optimization, 
 we can use Latin hypercube designs to select an initial set of parameters and then proceed to maximize the likelihood of the hyper-parameters iteratively \citep{Ye1998,Santner:2003}. This is the approach followed in our experiments. However, a good alternative is to use either classical or Bayesian quadrature to integrate out the hyper-parameters \citep{Osborne2010}.

Let $\mathbf{z}_{1:i}\sim \mathcal{N}(0,\mathbf{K})$ be the $i$ noisy
observations of the objective function obtained from previous
iterations. 
(Note that the Markov chain is run for $L$ steps for each
discrete iteration $i$. The extra index to indicate this fact has been
made implicit to improve readability.) $\mathbf{z}_{1:i}$ and $h_{i+1}$
are jointly multivariate Gaussian:
\begin{align*}
\begin{bmatrix}
\mathbf{z}_{1:i} \\
h_{i+1}
\end{bmatrix}
&=
\mathcal{N}
\left(
\mathbf{0},
\begin{bmatrix}
\mathbf{K}+\sigma^2_{\eta}I & \mathbf{k}^T \\
\mathbf{k} & k(\theta, \theta)
\end{bmatrix}
\right),
\end{align*}
where
\begin{align*}
\mathbf{K} &=
\begin{bmatrix}
k(\theta_1, \theta_1) & \ldots & k(\theta_1, \theta_i) \\
\vdots & \ddots & \vdots \\
k(\theta_i, \theta_1) & \ldots & k(\theta_i, \theta_i)
\end{bmatrix}
\end{align*}
and $\mathbf{k} =
[
k(\theta, \theta_1) \; \ldots \;
k(\theta, \theta_i)]^T.$ 
All the above assumptions about the form of the prior distribution and
observation model are standard and less restrictive than they might
appear at first sight. The central assumption is that the objective function is
smooth. For objective functions with discontinuities, we need more sophisticated surrogate functions for the cost. We refer readers to \cite{Gramacy:2004} and \cite{bayesopt_tut} for examples.

The predictive distribution for any value $\theta$ follows from the
Sherman-Morrison-Woodbury formula, where $\mathcal{D}_{1:i} =
(\theta_{1:i},\mathbf{z}_{1:i})$:
\begin{align*}
p(h_{i+1} | \mathcal{D}_{1:i},\theta) &=
\mathcal{N}
(\mu_i(\theta), \sigma^2_i(\theta)) \\
\mu_i(\theta) &=
\mathbf{k}^{T}(\mathbf{K + \sigma_{\eta}^2 I})^{-1}\mathbf{z}_{1:i} \\
\sigma^2_i(\theta) &=
k(\theta, \theta) -
\mathbf{k}^{T}(\mathbf{K + \sigma_{\eta}^2 I})^{-1}\mathbf{k}
\end{align*}

The next query point $\theta_{i+1}$ is chosen to maximize an
acquisition function, $u(\theta |\mathcal{D}_{1:i})$, that trades-off exploration (where
$\sigma^2_i(\theta)$ is large) and exploitation (where $\mu_i(\theta)$
is high). We adopt the expected improvement over the best candidate as
this acquisition function \cite{Schonlau:1998,bayesopt_tut}. This is a standard acquisition function for which asymptotic rates of convergence have been proved \citep{bull2011convergence}. However, we point out that there are a few other reasonable alternatives, such as Thompson sampling \citep{May2011} and upper confidence bounds (UCB) on regret \citep{Srinivas:2010}. A comparison among these options as well as portfolio strategies to combine them appeared recently in \citep{Hoffman2011}. There are several good ways of optimizing the acquisition function, including the method of DIvided RECTangles (DIRECT) of \cite{Finkel:2003} and many versions of the projected Newton methods of \cite{Bertsekas1987}. We found DIRECT to provide a very efficient solution in our domain. Note that optimizing the acquisition function is much easier than optimizing the original objective function. This is because the acquisition functions can be easily evaluated and differentiated.

\subsection{Sampling Phase}
\label{sec:sampling_phase}

The Bayesian optimization phase results in a Gaussian process on the $I$
noisy observations of the performance criterion $\mathbf{z}_{1:I}$,
taken at the corresponding locations in parameter space $\theta_{1:I}$.
This Gaussian process is
used to construct a discrete stochastic policy $p(\theta |
\mathbf{z}_{1:I})$ over the parameter space $\mbs{\Theta}$.
The Markov chain is run with
parameter settings randomly drawn from this policy at each step.

One can synthesize the policy $p(\theta|\mathbf{z}_{1:I})$ in several ways. The simplest is to use the mean of the GP to construct a distribution proportional to $\exp(\mu(\theta))$. This is the so-called Boltzmann policy.
We can sample $M$ parameter candidates $\theta_i$ according to this distribution. Our final sampler then consists of a mixture of $M$ transition kernels, where each kernel is parameterized by one of the $\theta_i$, $i=1,\ldots,M$. The distribution of the
samples generated in the sampling phase will approach the target
distribution $\pi(\cdot)$ as the number of iterations tends to
$\infty$ provided the kernels in this finite mixture are ergodic.

In high dimensions, one reasonable approach would be to use a multi-start optimizer to find maxima of the unnormalized Boltzmann policy and then perform local exploration of the modes with a simple Metropolis algorithm. This is a slight more sophisticated version of what is often referred to as the epsilon greedy policy.

The strategies discussed thus far do not take into account the uncertainty of the GP. A solution is to draw M functions according to the GP and then find he optimizer $\theta_i$ of each of these functions. This is the strategy followed in  \citep{May2011} for the case of contextual bandits. Although this strategy works well for low dimensions, it is not clear how it can be easily scaled.

\subsection{Objective Function}
\label{sec:perf-crit}

The auto-correlation $r(l, \mathcal{X})$ of a sequence of
$n$ generated samples $\mathcal{X} = \{\x^{(1)}, \ldots, \x^{(n)}\}$
as a function of the time lag between them is defined as
\begin{align*}
r(l, \mathcal{X}) &\triangleq
\frac{1}{(n-l)\delta^2}
\sum_{t=1}^{n-l}
(\x^{(t)} - \bar{\x})^{T}(\x^{(t+l)} - \bar{\x}),
\end{align*}
where $l$ is the lag and $\bar{\x}$ and $\delta^2$ are the mean and
the variance of $\mathcal{X}$, respectively.

Faster mixing times are characterized by larger values of 
$a(l_{\textrm{max}}, \mathcal{X}) = 1- 
({l_{\textrm{max}}}^{-1}) \sum_{l=1}^{l_{\textrm{max}}}
|r(l, \mathcal{X})|$. We use this property to construct the criterion for Bayesian optimization as follows.
Let $\mathcal{E}_{i}$ be the last $i$
sampled states (the energies in the experimental section).
The performance criterion is
obtained by taking the average of $a(\cdot)$ over a
sliding window within the last $L$ samples, down to a minimum window
size of $25$: $\frac{1}{L-25+1} \sum_{i=25}^{L}
a(i,\mathcal{E}_{i})$.


\section{Application to Constrained Discrete Distributions}
\label{sec:constrained}

The Intracluster Move (IM) sampler was recently
proposed to generate samples from notoriously-hard constrained
Boltzmann machines in \citep{im_expert}. This sampler has two
parameters (one continuous and the other discrete) that the authors
state to be difficult to tune. This and the recent growing interest in discrete Boltzmann machines in machine learning motivated us
to apply the proposed Bayesian-optimized MCMC method to this problem.

Boltzmann machines are described in \cite{ackley85alearning}. Let $x_i
\in \{0,1\}$ denote the $i$-th random variable in the model. The
Boltzmann distribution is given by
\begin{align}
\pi(\state{x}) &\define \frac{1}{Z(\beta)} \energyfn{\state{x}},
\end{align}
where $Z(\beta) \define \sum_{\state{x} \in \statespace}
\energyfn{\state{x}}$ is the normalizing constant, $\beta$ is a
temperature parameter and $E({\state{x}}) \define - \sum_{i,j} x_i
J_{ij} x_{j} - \sum_{i} b_i x_i$ is the energy function.
Boltzmann machines also have coupling parameters $J$ and $b$ that are
assumed to be known.

Let $\statespace_n(\refstate)$ be the subset of the states that are at
exactly Hamming distance $n$ away from a reference state
$\refstate$. The distribution $\pi_{n, \refstate}(\state{x})$ is the
restriction of $\pi(\state{x})$ to
$\statespace_n(\refstate)$. $\pi_{n, \refstate}(\state{x})$ has
\mbox{$Z_n(\beta, \refstate) \define \sum_{\state{x} \in
    \statespace_n(\refstate)} \energyfn{\state{x}}$} as its
normalizing constant and is defined as
\begin{align}
\pi_{n, \refstate}(\state{x}) &\define
\left\{
\begin{array}{ll}
\frac{1}{Z_n(\beta, \refstate)} \energyfn{\state{x}} &
\text{if } \state{x} \in \statespace_n(\refstate) \\
0 & \text{otherwise}
\end{array}
\right.
\end{align}
The rest of the paper makes $\refstate$ implicit and uses the
simplified notation $\statespace_n$, $\pi_n(\x)$ and $Z_n(\beta)$.
These constraints on the states are used in statistical physics and in regularized
statistical models \citep{im_expert}.

The IM sampler proposes a new state $\y^{(t+1)} \in \statespace_n$ from
an original state $\x^{(t)} \in \statespace_n$ using self-avoiding
walks (SAWs) and has parameters $\theta = (k, \gamma)$, where $k \in \mathcal{L} \triangleq \{1,2,\ldots,k_{\max}\}$ is the length of each SAW and 
$\gamma \in \mathcal{G} \triangleq [0,\gamma_{\max}]$ is the
energy-biasing parameter. $k$ determines the size, in terms of the
number of bits flipped, of the moves through $\statespace_n$. $\gamma$
controls the degree to which higher energy states are favored.

\subsection{Experimental Setup}

The experiments compare the performance of four different sampling
methods on three different models.
The sampling methods are all instances of the IM sampler that only
differ in the manner that $\gamma$ and $k$ are picked:
\begin{description}
\setlength{\itemsep}{2pt}
\item[Kawasaki sampler]: It transitions from state to
 state within $\statespace_n$ by uniformly sampling a bit to flip to
  produce a state in $\statespace_{n+1}$ or $\statespace_{n-1}$ and
  then uniformly sampling a bit to flip to return to
  $\statespace_n$ \citep{Kawasaki1966}. This is equivalent to running the IM sampler with
  $\gamma$ fixed to 0 and $k$ fixed to 1.
\item[IMExpert] is the IM sampler manually tuned by a domain expert \citep{im_expert}. The expert set $\gamma$ to a fixed value
  $\gamma_{\textrm{expert}}$ and $k$ is drawn uniformly in the set $\mathcal{L}$.
\item[IMUnif] is a completely naive approach that draws $\gamma$
  uniformly from $\mathcal{G}$ and $k$ uniformly from $\mathcal{L}$.
\item[IMBayesOpt] is the Bayesian-optimized IM with ${L = 100}$ samples
  generated for each of the $100$ adaptations of the parameters.
\end{description}

The algorithm parameters for each model, defined subsequently, are shown in Table
\ref{table:algo_params}, where ``Others'' refers to IMUnif and
IMBayesOpt. These two samplers have the same parameter sets because
IMUnif is a baseline algorithm used to ensure that Bayesian-optimized
IM performs better than a naive strategy.
\begin{table}[t]
\caption{Algorithm parameters and sets for each model following \citep{im_expert}}
\label{table:algo_params}
\begin{center}
\begin{tabular}{llll}
\textbf{MODEL} & \textbf{ALGORITHM }& $\sawlenset$ & $\gammaset$\\
\hline \\
2DGrid & IMExpert & $\{90\}$ & $\{0.44\}$ \\
2DGrid & Others & $\{1, \ldots, 300\}$ & $[0, 0.88]$ \\
3DCube & IMExpert & $\{1, \ldots, 25\}$ & $\{0.8\}$ \\
3DCube & Others & $\{1, \ldots, 50\}$ & $[0, 1.6]$ \\
RBM & IMExpert & $\{1, \ldots, 20\}$ & $\{0.8\}$ \\
RBM & Others & $\{1, \ldots, 50\}$ & $[0, 1.6]$ \\
\end{tabular}
\end{center}
\end{table}
The parameters for IMUnif and IMBayesOpt were selected such that
$\sawlenset$ is a much larger superset of the SAW lengths used for
IMExpert and $\gammaset$ is the contiguous interval from $0$ to
$2\gamma_{\textrm{expert}}$. The parameters for IMExpert come from
\citep{im_expert}. The Kawasaki sampler does not have any parameters.

\begin{table}[h!]
\caption{Model parameters from \cite{im_expert}.  $n$ refers to the
   number of bits that
  are set to $1$ out of the total number of bits. $\beta^{-1}$ is the
  temperature.}
\label{table:model_params}
\begin{center}
\begin{tabular}{llll}
\textbf{MODEL} & $\beta^{-1}$ & \textbf{SIZE} & $n$ \\
\hline \\
2DGrid & $2.27$ & $60 \times 60$ & $1800$ of $3600$ \\
3DCube & $1.0$ & $9 \times 9 \times 9$ & $364$ of $729$ \\
RBM & $1.0$ & $|v| = 784$, $|h| = 500$ & $428$ of $1284$ \\
\end{tabular}
\end{center}
\end{table}

We consider the three models studied by \cite{im_expert}.
The model
parameters are given in Table \ref{table:model_params}. Note that $n$
refers to the Hamming distance from states in $\statespace_n$ to the
reference state $\state{c}$. The reference state $\state{c}$ was set to the ground
state where none of the bits are set to $1$. This is particularly intuitive in machine learning applications, where we might not want more than a small percentage of the binary variables on as part of a regularization or variable selection strategy.

\paragraph{Ferromagnetic 2D grid Ising model:}
The ferromagnetic 2D grid Ising model is made up of nodes arranged in
a planar and rectangular grid with connections between the nodes on
one boundary to the nodes on the other boundary for each dimension
(i.e. periodic boundaries), also known as a square toroidal
grid. Hence, each node has exactly four neighbours. The interaction
weights, $J_{ij}$, are all 1 and the biases, $b_i$, are all 0. The temperature 2.27 is
the critical temperature of the model.

\paragraph{Frustrated 3D cube Ising model:}
The frustrated 3D cube Ising model is made up of nodes arranged in a
topology that is the three-dimensional analogue of the two-dimensional
grid with periodic boundaries. Hence,
each node has exactly six neighbours. The interaction weights,
$J_{ij}$, are uniformly sampled from the set $\{-1, 1\}$ and the
biases, $b_i$, are all 0.

\paragraph{Restricted Boltzmann machine (RBM):}
The RBM has a bipartite graph structure, with $h$ hidden nodes in
one partition and $v$ visible nodes in the other.  The interaction
weights $J_{ij}$ and biases $b_i$ are exactly the same as in
\cite{im_expert} and correspond to local Gabor-like filters that capture
regularities in perceptual inputs.

Each sampler was run five times with $9 \times 10^4$ steps for each
run. IMBayesOpt had an additional $10^4$ steps for an adaptation phase
consisting of $100$ adaptations of $100$ samples each. IMBayesOpt was
not penalized for the computational overhead involved in these
additional steps because it is seen as being far cheaper than having
the IM sampler parameters tuned manually.

\begin{figure}[h]
  \begin{tabular*}{10 cm}{c c}
    \includegraphics[width=0.49\columnwidth]{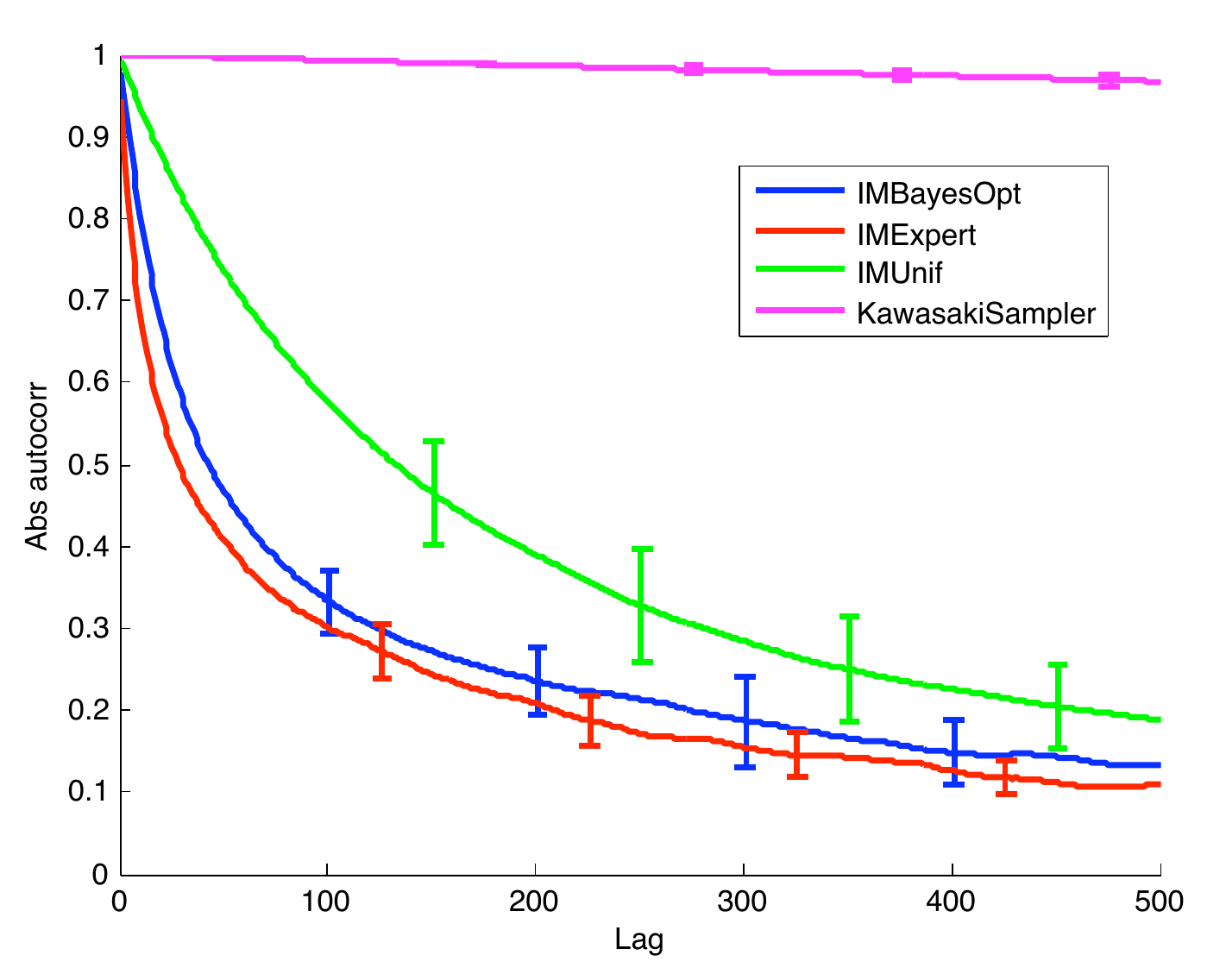} &
    \includegraphics[width=0.49\columnwidth]{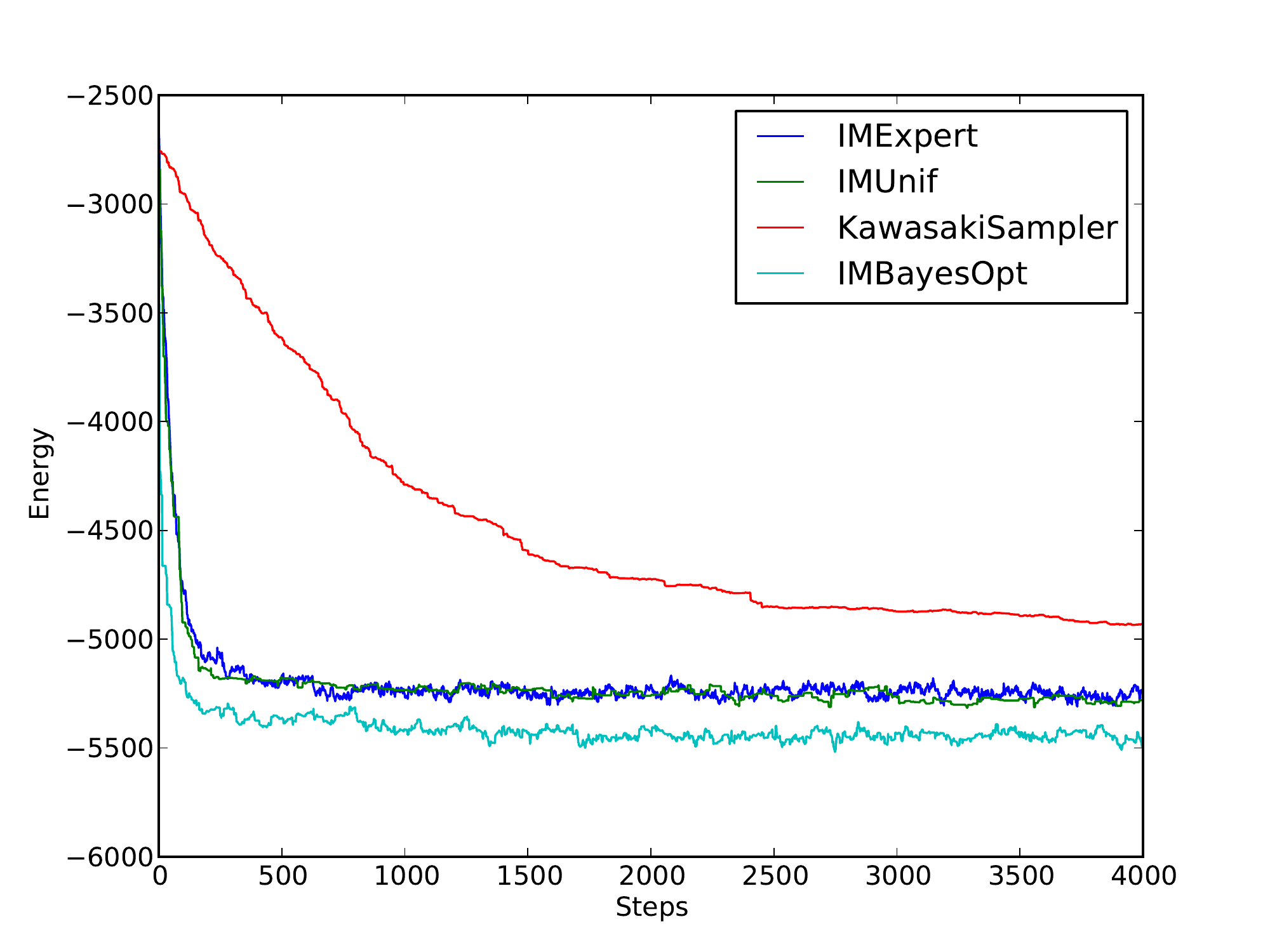} \\

  \end{tabular*}
  \caption{[Left] Mean and error bars of the auto-correlation function
  of the energies of the sampled states drawn from the 2D grid Ising model. [Right] Average of the energies.}
  \label{fig:ferro2d_acf_compare}
\end{figure}

\begin{figure}[h]
\centering
\includegraphics[width=0.49\columnwidth]{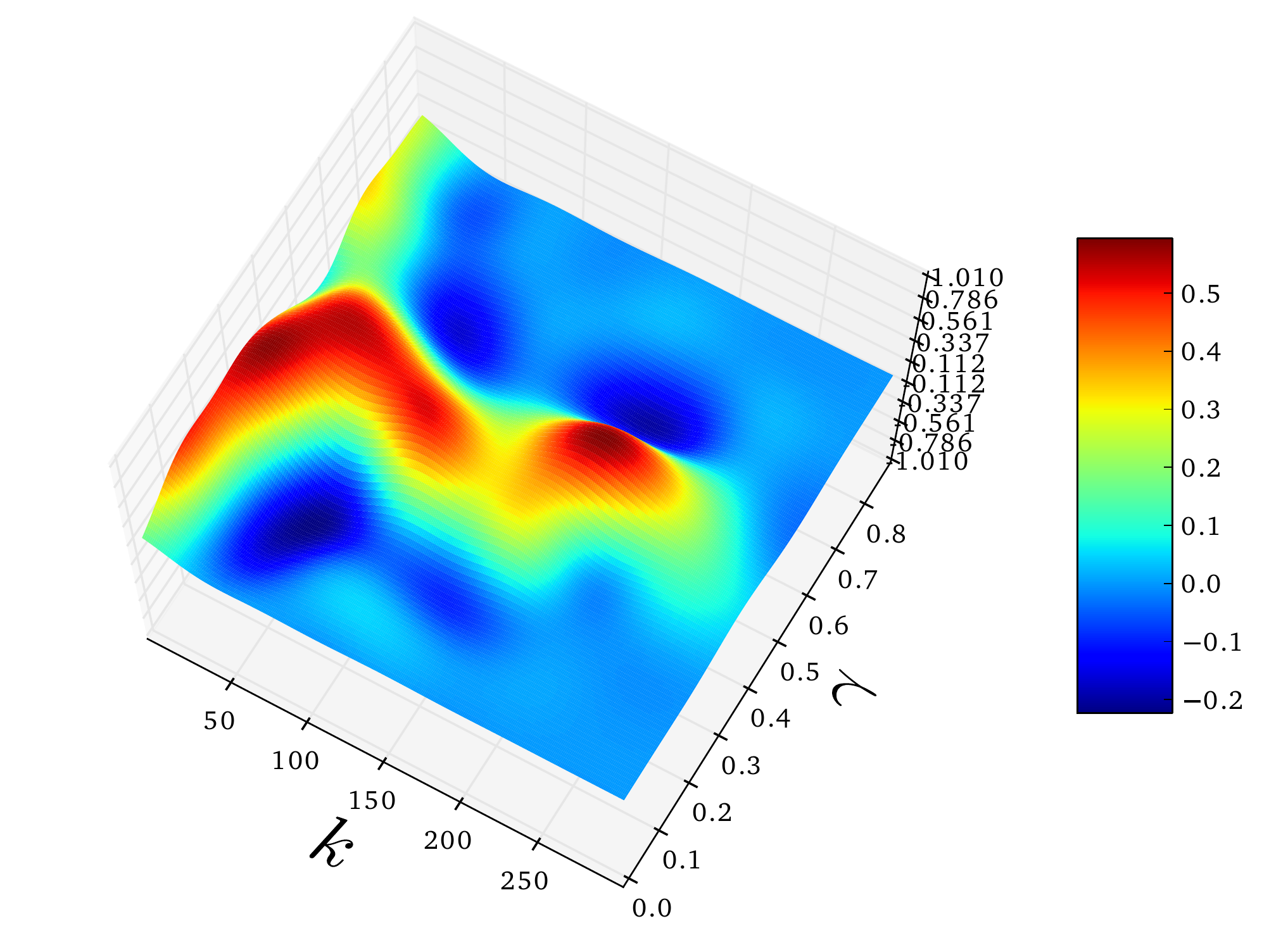}
\includegraphics[width=0.49\columnwidth]{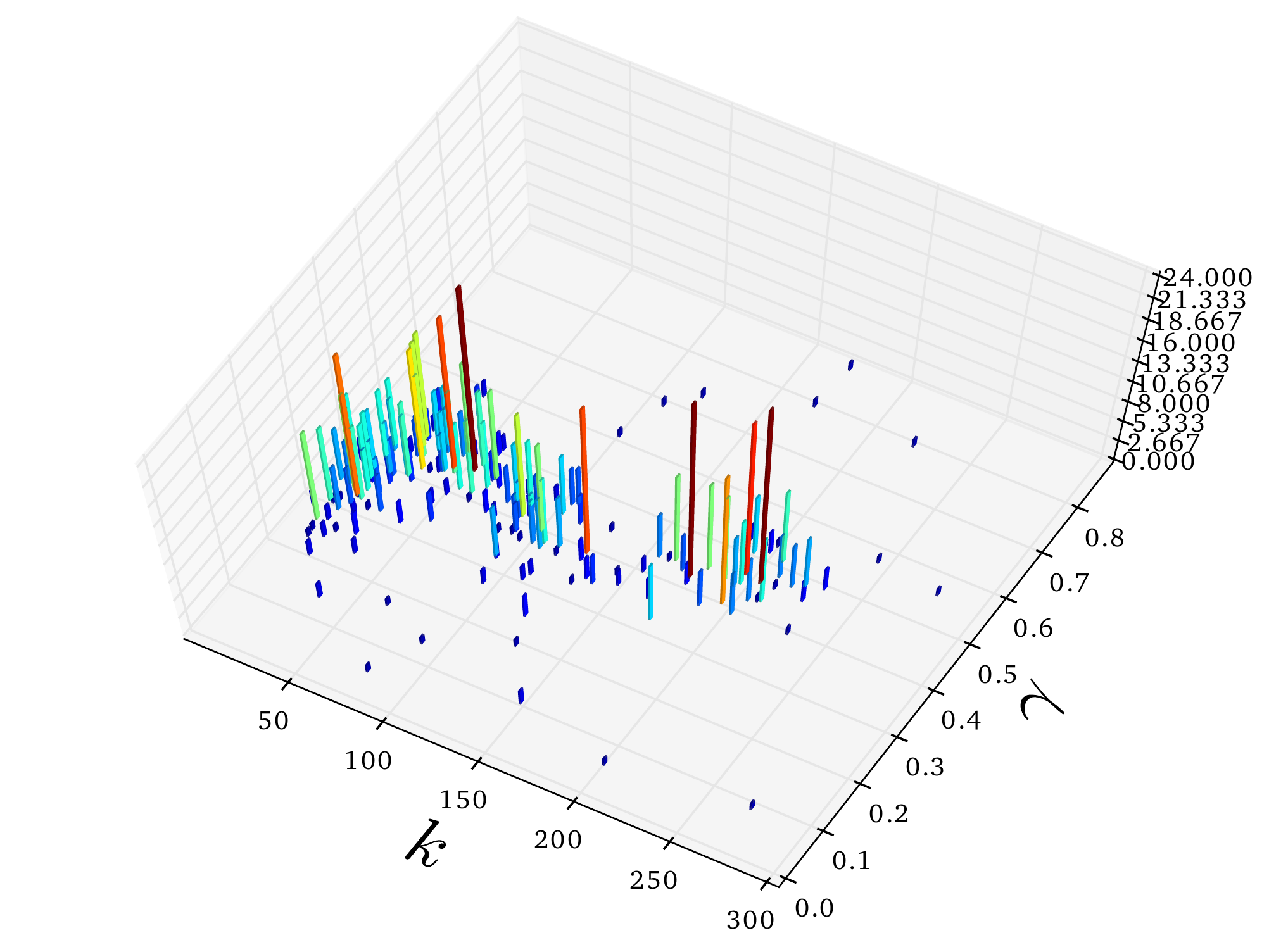}
\caption{[Left] The mean function of the Gaussian processes over
  $\mbs{\Theta}$, learned by IMBayesOpt for the 2D grid Ising
  model. An average over the five trials is shown. [Right] The corresponding policy used for sampling.}
\label{fig:ferro2d_im_ego_perf_gp_mean}
\end{figure}

All of the algorithms have a burn-in phase consisting of the first
$10^4$ samples generated in the corresponding sampling phase. The
burn-in phase was not included in the computation of the
auto-correlation functions in figures \ref{fig:ferro2d_acf_compare},
\ref{fig:isg3d_E_samples} and \ref{fig:rbm_E_samples}.

IMBayesOpt begins its sampling phase in the same starting state as all
of the other samplers, even though it would most likely be in a low
energy state at the end of its adaptation phase. This ensures that the
comparison of the sampling methods is fair.

\subsection{Results}
\label{sec:results}

\subsubsection{Ferromagnetic 2D grid Ising model}
\label{sec:ferro2d}

IMExpert and IMBayesOpt both have very similar mean ACFs, as indicated in figure
\ref{fig:ferro2d_acf_compare}. IMUnif suffers from many long strings of consecutive
proposal rejections. This is evident from the many intervals where the
sampled state energy does not change.

%

\begin{figure}[h]
  \begin{tabular*}{10 cm}{c c}
    \includegraphics[width=0.49\columnwidth]{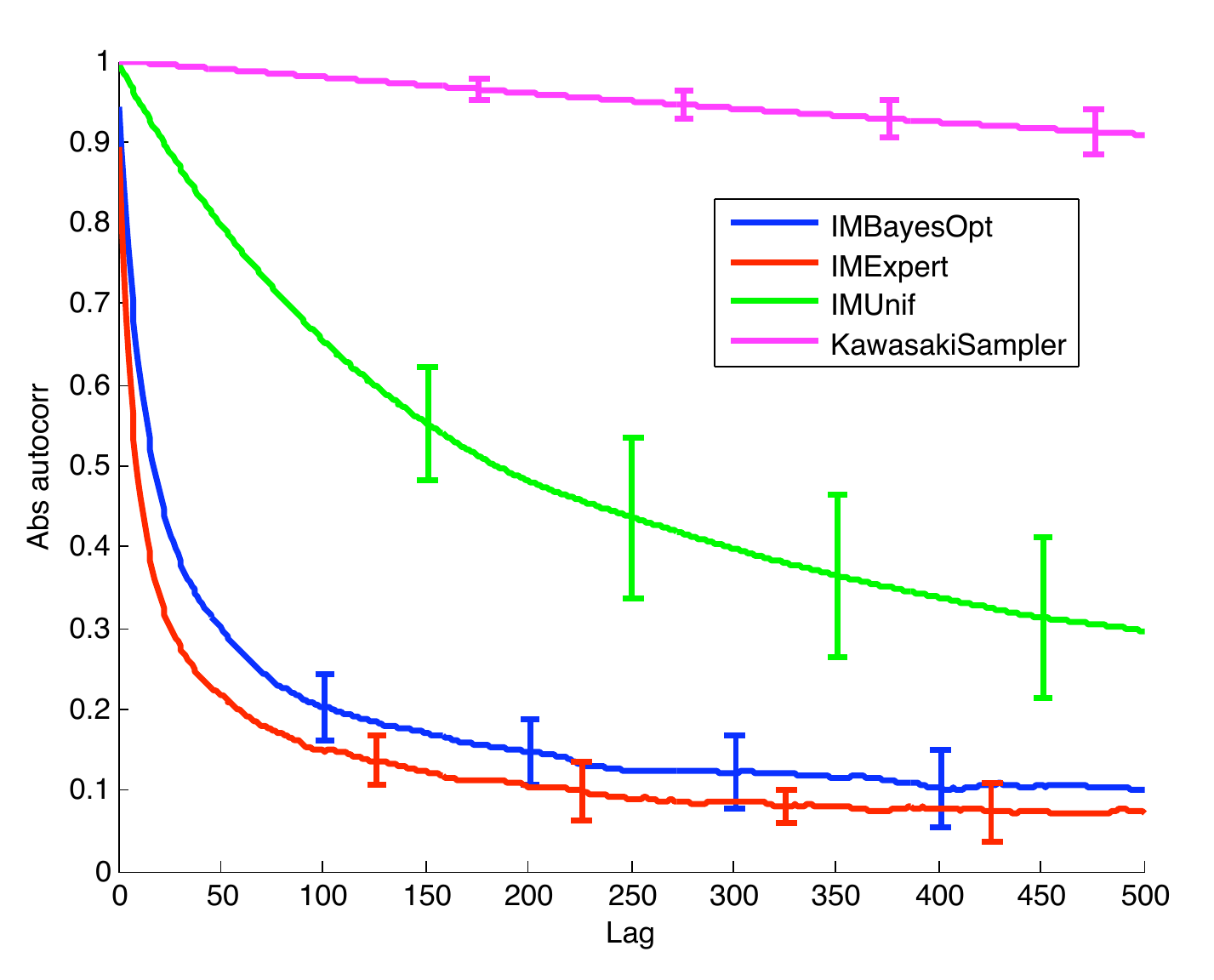} &
    \includegraphics[width=0.49\columnwidth]{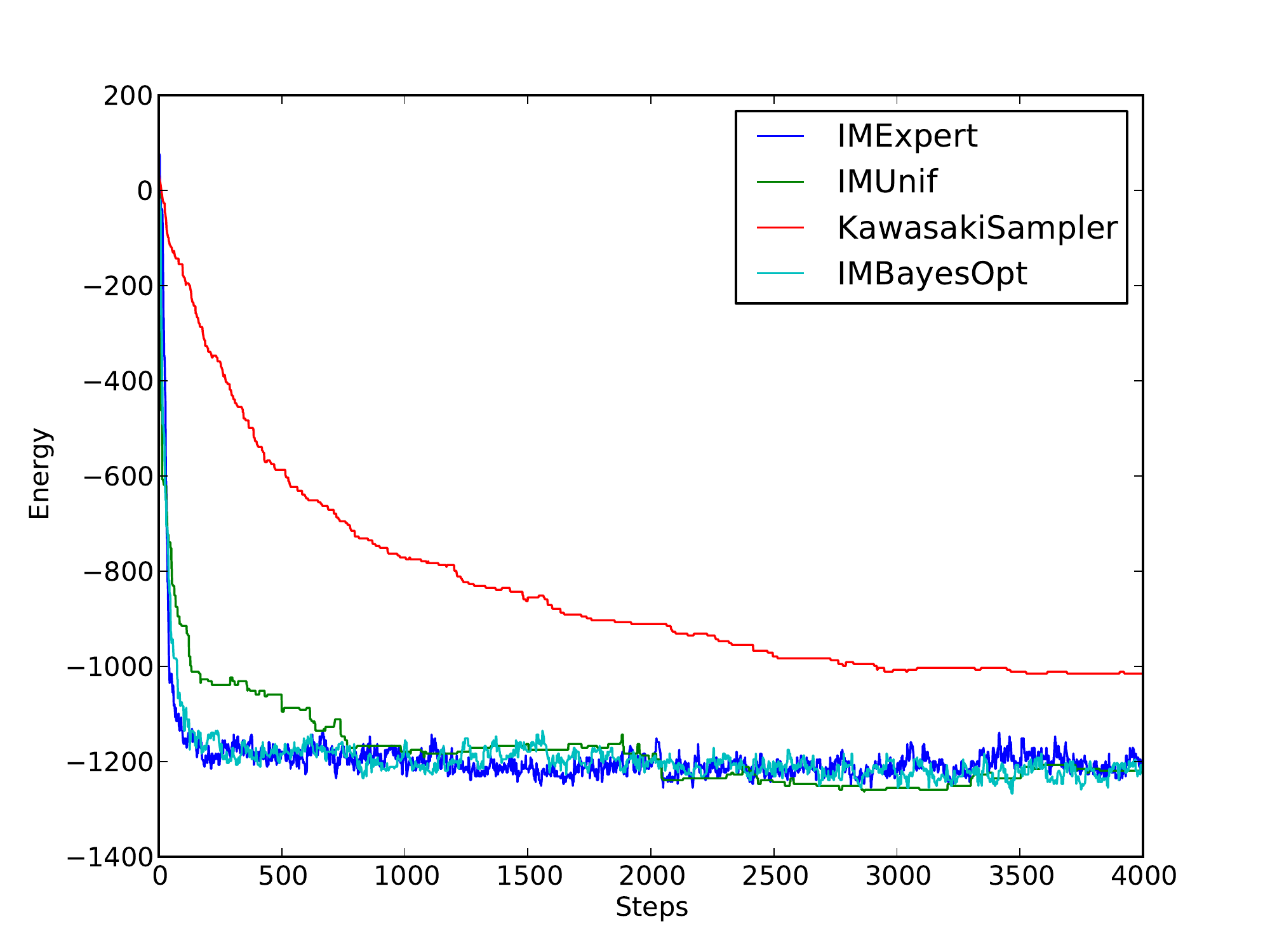} \\

  \end{tabular*}
  \caption{[Left] Mean and error bars of the auto-correlation function
  of the energies of the sampled states drawn from the 3D cube Ising
  model. [Right] Average of the energies.}
  \label{fig:isg3d_E_samples}
\end{figure}

Figure \ref{fig:ferro2d_im_ego_perf_gp_mean} suggests that $\gamma$ is
much more important to the performance of the IM sampler than the SAW
lengths for this model, especially at large SAW lengths. One of the highest probability points in
the Gaussian process corresponds to the parameters chosen by
\cite{im_expert} $(\gamma = 0.44, k = 90)$. Figure \ref{fig:ferro2d_im_ego_perf_gp_mean} also shows $M=1000$ samples drawn from the Boltzmann policy. The MCMC results are for this policy. The same procedure is adopted for the other models.

\subsubsection{Frustrated 3D cube Ising model}
\label{sec:isg3d}

Figure \ref{fig:isg3d_E_samples} shows
that IMUnif now performs far worse than the IMExpert and IMBayesOpt,
implying that the extremely rugged energy landscape of the 3D cube
Ising model makes manual tuning a non-trivial and necessary
process. IMBayesOpt performs very similarly to IMExpert, but is
automatically tuned without any human intervention.

\begin{figure}[h]
\centering
\includegraphics[width=0.49\columnwidth]{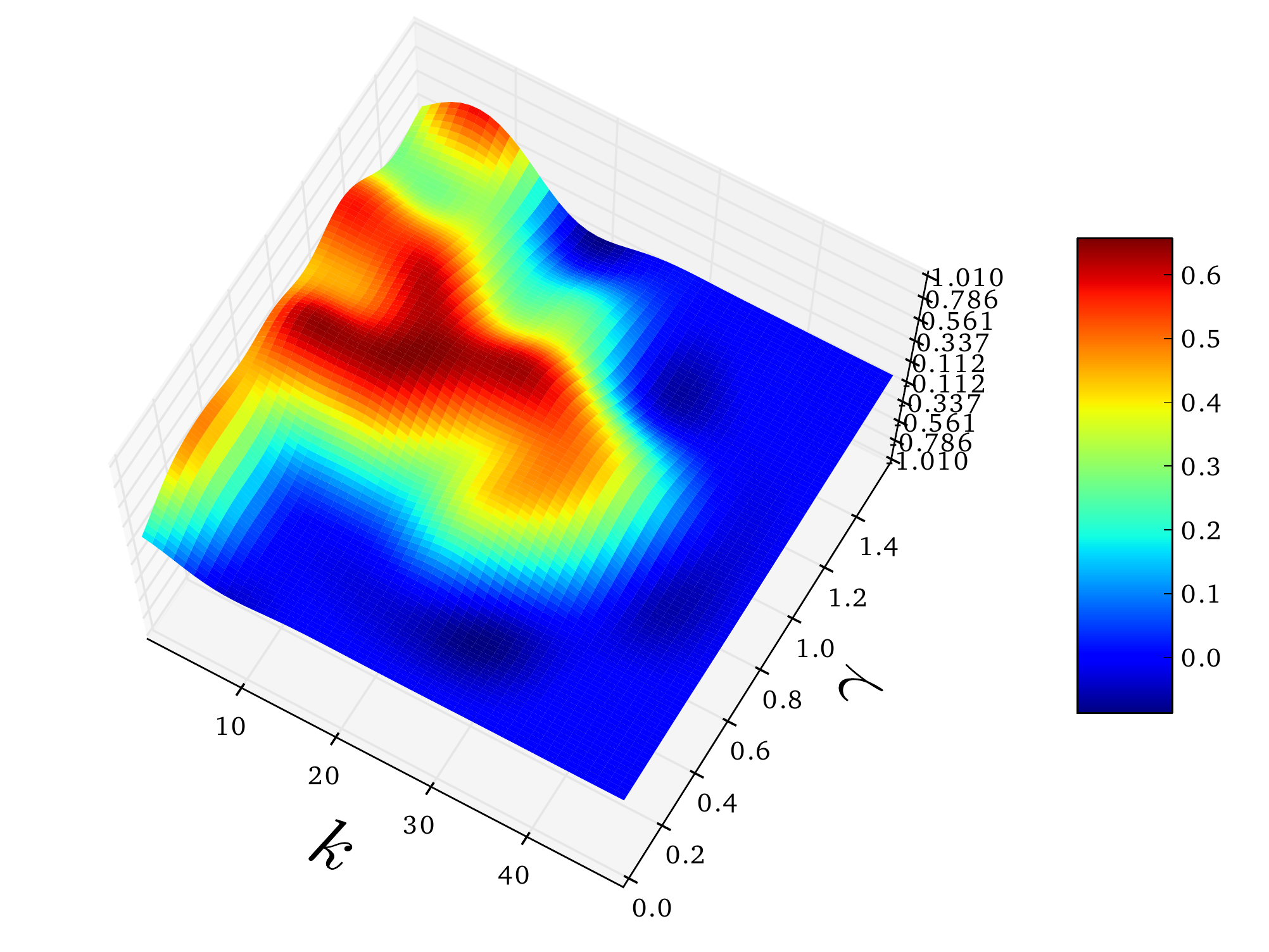}
\includegraphics[width=0.49\columnwidth]{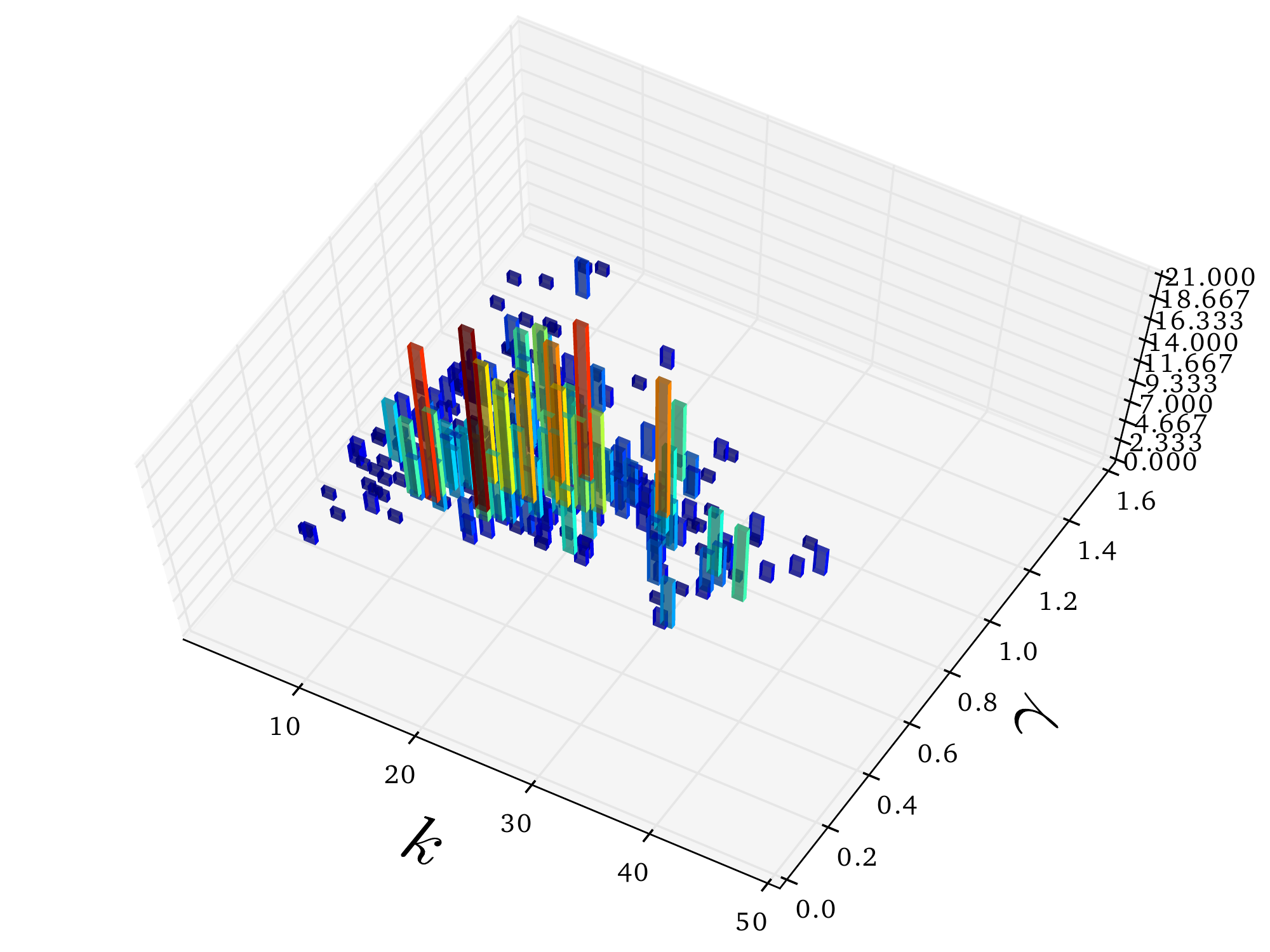}
\caption{[Left] The mean function of the Gaussian processes over
  $\mbs{\Theta}$, learned by IMBayesOpt for the 3D cube Ising
  model. An average over the five trials is shown. [Right] The corresponding policy used for sampling.}
\label{fig:isg3d_im_ego_perf_gp_mean}
\end{figure}

The Gaussian process mean function in Figure
\ref{fig:isg3d_im_ego_perf_gp_mean} suggests that SAW lengths should
not be longer than $k = 25$, as found in \cite{im_expert}. Both
IMExpert and IMBayesOpt are essentially following the same strategy
and performing well, while the performance of IMUnif confirms that
tuning is important for the 3D cube Ising model.

\subsubsection{Restricted Boltzmann Machine (RBM)}
\label{sec:rbm}


\begin{figure}[h]
  \begin{tabular*}{10 cm}{c c}
    \includegraphics[width=0.49\columnwidth]{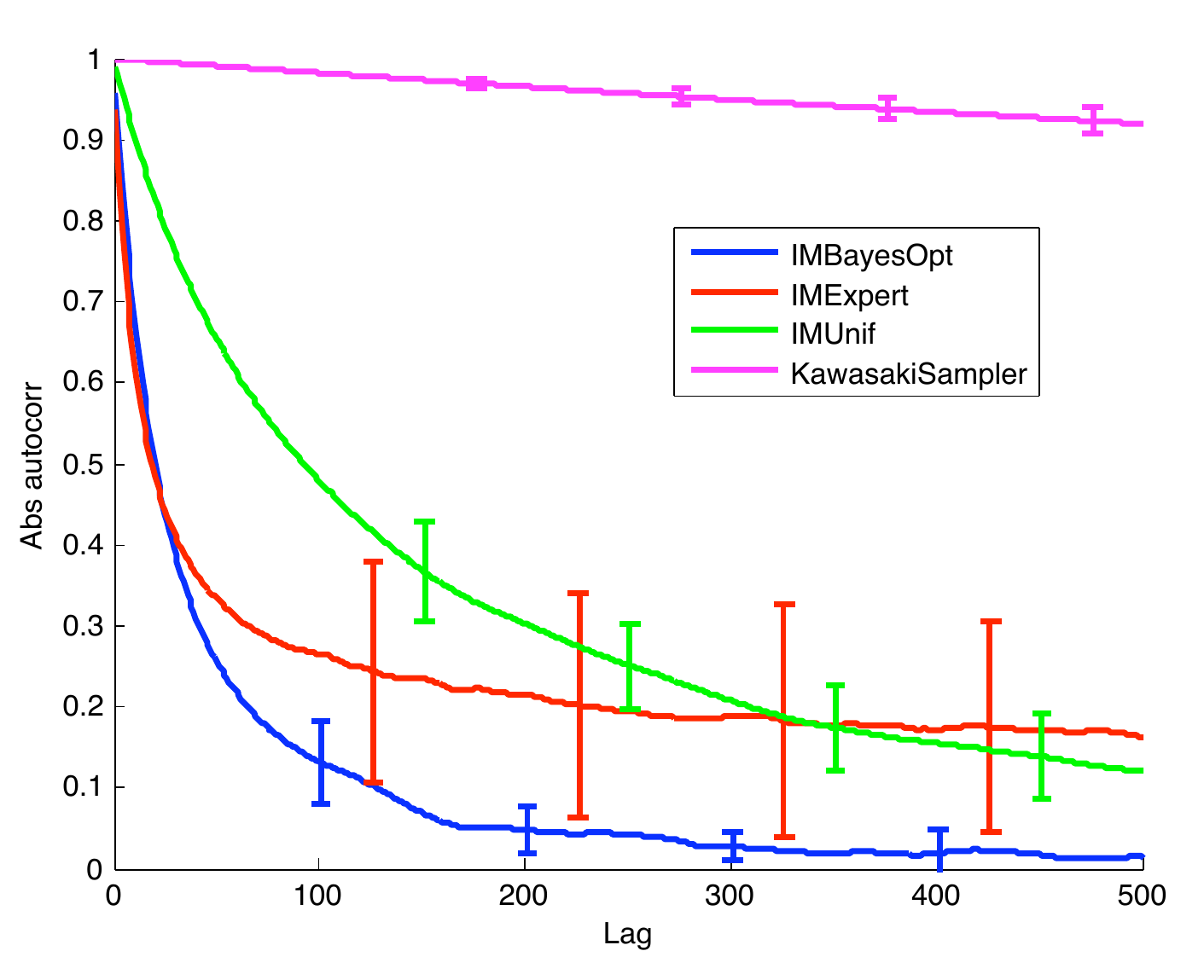} &
    \includegraphics[width=0.49\columnwidth]{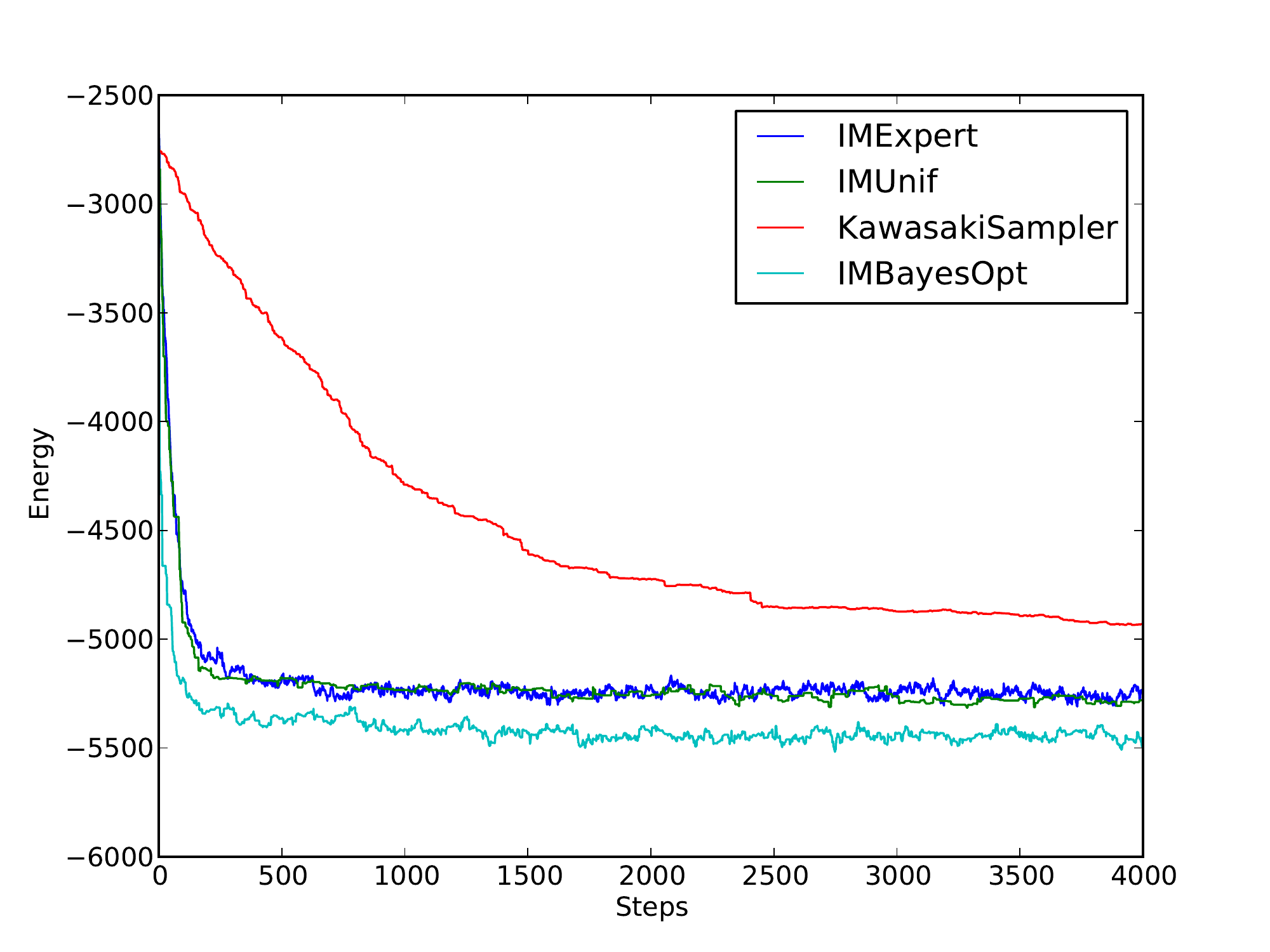} \\

  \end{tabular*}
  \caption{[Left] Mean and error bars of the auto-correlation function
  of the energies of the sampled states drawn from the 3D cube Ising
  model. [Right] Average of the energies.}
  \label{fig:rbm_E_samples}
\end{figure}

\cite{im_expert} find that IMExpert experiences a rapid dropoff in
ACF, but this dropoff is exaggerated by the inclusion of the burn-in
phase. Figure \ref{fig:rbm_E_samples} shows a much more modest
dropoff when the burn-in phase is left out of the ACF
computation. However, it still
corroborates the claim in \cite{im_expert} that IMExpert performs far
better than the Kawasaki sampler.

\begin{figure}[h]
\centering
\includegraphics[width=0.49\columnwidth]{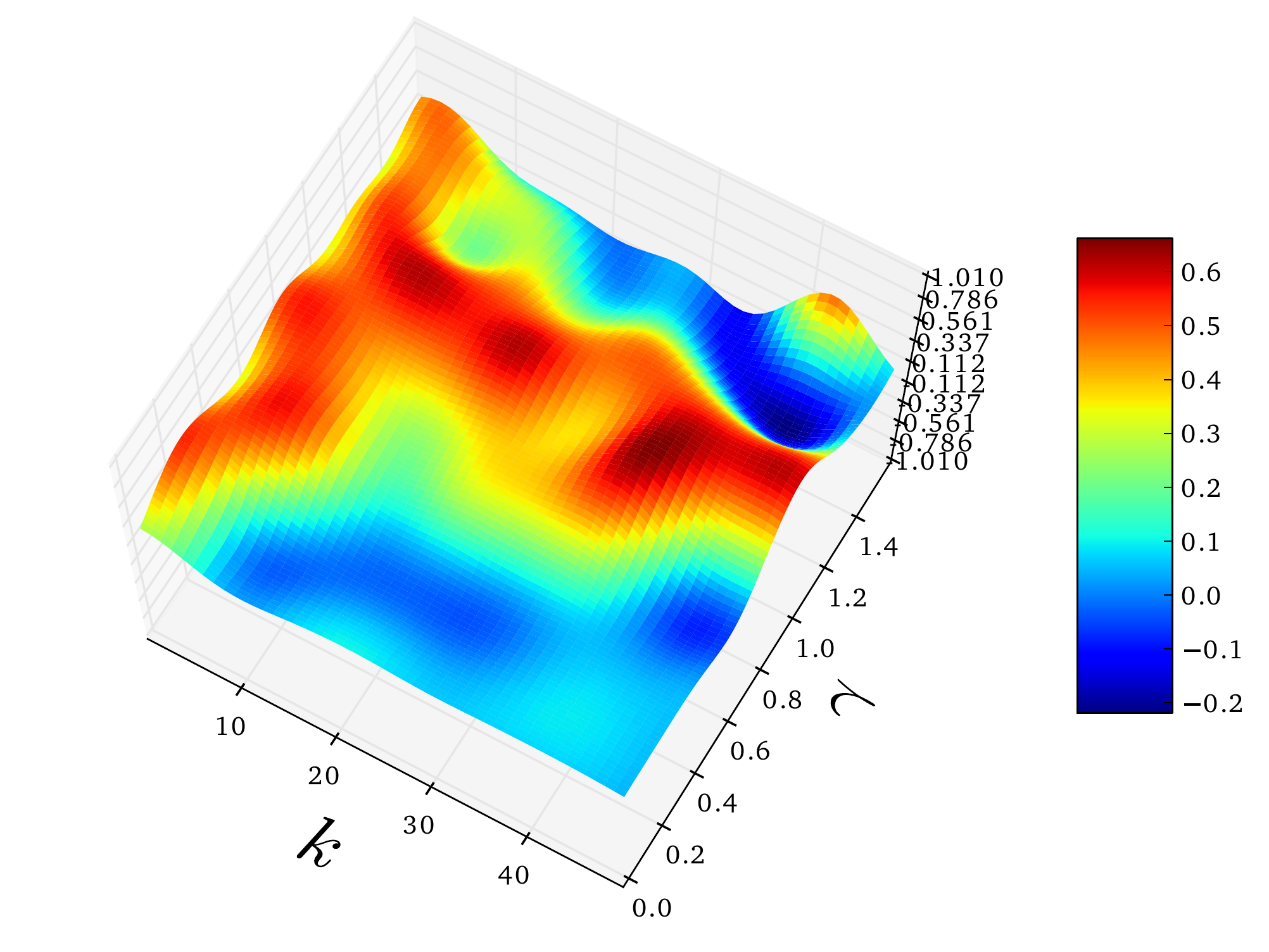}
\includegraphics[width=0.49\columnwidth]{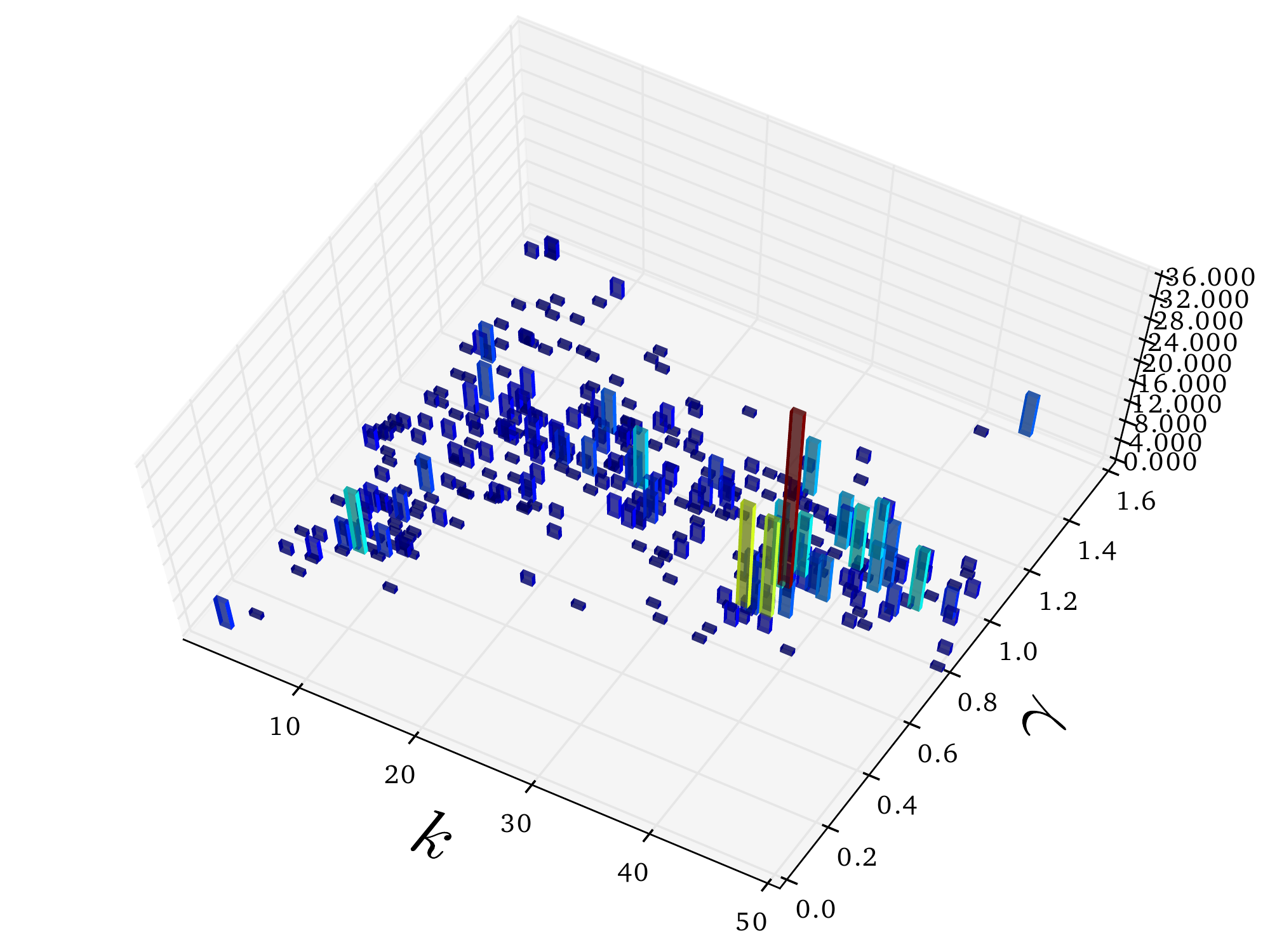}
\caption{[Left] The mean function of the Gaussian processes over
  $\mbs{\Theta}$, learned by IMBayesOpt for the RBM. An
  average over the five trials is shown. [Right] The corresponding policy used for sampling.}
\label{fig:rbm_im_ego_perf_gp_mean}
\end{figure}

Figure \ref{fig:rbm_E_samples} 
shows that IMExpert does not perform much better than IMUnif. The
variance of IMExpert's ACF is also much higher than any of the other
methods. IMBayesOpt performs significantly better than any of the other methods,
including manual tuning by a domain expert.

The Gaussian process mean function in figure
\ref{fig:rbm_im_ego_perf_gp_mean} suggests that SAW lengths greater
than 20 can be used and are as effective as shorter ones, whereas
\cite{im_expert} only pick SAW lengths between $1$ and $20$. This
discrepancy is an instance where Bayesian-optimized MCMC has found
better parameters than a domain expert.


\section{Conclusions}
Experiments were conducted to assess Bayesian optimization for
adaptive MCMC. \cite{im_expert} state that it is easiest to manually
tune the 2D grid Ising model and that tuning the 3D cube Ising model
is more challenging. Manual tuning significantly improves the
performance of the IM sampler for these cases, but Bayesian-optimized
MCMC is able to realize the same gains without any human
intervention. This automatic approach surpasses the human expert for the RBM model.

Bayesian optimization is a general method for adapting the
parameters of any MCMC algorithm. It has some advantages over
stochastic approximation, as discussed in this
paper. However, it presently only applies to parameter spaces of up to
fifty dimensions. 
Bayesian optimization should not be seen as a replacement for stochastic approximation,
but rather as a complementary technique. In particular, Bayesian optimization should be adopted when the objective is non-differentiable or expensive to evaluate.
%
%


{
\bibliography{adaptsaw,thesis,mc}
\bibliographystyle{icml2011}
}

\end{document}